\documentclass[fleqn,twoside]{article}
\usepackage{espcrc2}

\usepackage{graphicx}
\usepackage[figuresright]{rotating}

\newcommand{\AmS}{{\protect\the\textfont2
  A\kern-.1667em\lower.5ex\hbox{M}\kern-.125emS}}

\newcommand{\lsim}{\mathrel{\mathop{\kern 0pt \rlap
  {\raise.2ex\hbox{$<$}}}
  \lower.9ex\hbox{\kern-.190em $\sim$}}}
\newcommand{\gsim}{\mathrel{\mathop{\kern 0pt \rlap
  {\raise.2ex\hbox{$>$}}}
  \lower.9ex\hbox{\kern-.190em $\sim$}}}

\def\ee{\end{equation}}
\def\be{\begin{equation}}
\def\eea{\end{eqnarray}}
\def\bea{\begin{eqnarray}}
\def\eeas{\end{eqnarray*}}
\def\beas{\begin{eqnarray*}}

\title{Supersymmetric Dark Matter}

\author{A. Bottino, F. Donato, N. Fornengo and S. Scopel
 \address[MCSD]{Dipartimento di Fisica Teorica, Universit\`a di Torino \\
Istituto Nazionale di Fisica Nucleare, Sezione di Torino \\
via P. Giuria 1, I--10125 Torino, Italy \\
Web: http://www.to.infn.it/astropart}
\thanks{Presented by A. Bottino (e--mail: bottino@to.infn.it)}
}
       
\begin{document}

\begin{abstract}

After a short introduction on some general implications of recent cosmological
observations, we discuss the properties of relic neutralinos in
connection with present experimental strategies for detecting 
WIMPs.  
\vspace{1pc}
\end{abstract}

\maketitle

\section{Introduction}
\label{sec:intro}

A large number of independent astronomical observations (on galactic
halos, clusters of galaxies, large scale structures) indicate that in
our Universe the average density of matter, normalized to the critical
density $\rho_c = 1.88 \cdot 10^{-29} \; h^2 \; {\rm g} \cdot {\rm
cm}^{-3}$ ({\it h} is the Hubble constant in units of 100 km $\cdot
{\rm ~s}^{-1} \cdot {\rm Mpc}^{-1}$), is in the range $0.2 \lsim \Omega_m
\lsim 0.4$ (or equivalently, $0.05 \lsim \Omega_m h^2 \lsim 0.3$)
\cite{primack}.  The value $\Omega_m \sim 0.3$ may also be derived by
combining CMB measurements \cite{cmb} with data on high-redshift SNIa
\cite{sn}, with the further result that a large value,
$\Omega_{\Lambda} \sim 0.7$, should be assigned to the
cosmological-constant (or quintessence) contribution.

By comparing these results on $\Omega_m$ with the contribution to matter
provided by visible matter, $\Omega_{vis} \sim  0.003$, and with the 
amount of baryonic matter, as deduced from primordial nucleosynthesis, 
$\Omega_b \lsim  0.05$ \cite{bbn}, one concludes that: i) most of the matter 
in the Universe is dark, ii) only a small fraction of it is 
baryonic. 

Particle physics offers a large selection of possible candidates for
dark matter, once one considers extensions of the Standard Model
(SM). The most obvious option is represented by massive neutrinos, as
suggested by oscillation effects observed in solar and atmospheric
neutrinos (for the latest data, see Ref.\cite{solar} and
Ref.\cite{atm}, respectively). In these experiments only differences
in neutrino squared masses are measured; from these results and from
the standard formula $\Omega_{\nu} h^2 = \sum_{\nu} \; m_{\nu}/(93 \;
{\rm eV})$ a lower bound may be established for neutrino contribution
to the cosmological matter density, $\Omega_{\nu} h^2 \gsim 6 \cdot
10^{-4}$.  Obviously, the cosmological neutrino contribution may be
larger, in case of neutrino mass degeneracy. However, taking into
account the stringent upper bound, $m_{\nu_e} < 2.2$ eV, derived from
Tritium beta decay \cite{tritium}, an upper limit on the neutrino
cosmological contribution may be derived: $\Omega_{\nu} h^2 \lsim
0.07$ (or $\Omega_{\nu} \lsim 0.14$, using the central value $h =
0.7$).

The existence of an upper bound of roughly this order of magnitude for
the neutrino contribution to the matter density in the Universe is not
unexpected, since the evolution theory of the primordial density
fluctuations into the present cosmological structures indicates that
most of the dark matter must be comprised of cold particles, {\it
i. e.} of particles that decoupled from the primordial plasma when
nonrelativistic.  Massive neutrinos, with masses up to about 1 MeV,
are relativistic at decoupling, and then, by their free streaming,
tend to erase small--scale structures. A recent investigation
\cite{elgaroy}, based on the largest galaxy redshift survey (2dFGRS),
has derived the upper bound $\Omega_{\nu} \lsim 0.05$.

Thus, by using particle physics results and cosmological observations,
one arrives at the conclusion that light neutrinos are expected to
only constitute a subdominant dark matter component.

We turn now to cold relics.  The order of magnitude of the relic
abundance of these cold particles, or WIMPs (Weakly Interacting
Massive Particle), is set by the approximate formula $\Omega_{\chi}
h^2 \sim 3 \; \cdot 10^{-27} {\rm cm}^3 {\rm ~s}^{-1} / <\sigma_{a} \;
v>_{int}$, where $ <\sigma_{a} \; v>_{int}$ denotes the thermal
average of the product of the annihilation cross-section $\sigma_a$
times the relative velocity $v$ for a pair of WIMPs, integrated from
the freeze--out temperature $T_f$ to the present--day one
\cite{noiomega}.

As mentioned above, to identify a WIMP with a specific particle
candidate, one has to move beyond the Standard Model. The only
guidance in the selection of a specific model is provided by
theoretical motivations. Indeed, from measurements at accelerators and
in other precision experiments no significant evidence of physics
beyond the SM emerged so far; thus we can only count on bounds, which
significantly constrain the amount of acceptable deviations from the
SM predictions.

One of the most strongly motivated extension of the Standard Model is
represented by Supersymmetry. This is the theoretical framework within
which we consider here possible candidates for dark matter.

\section{Supersymmetric models}

Supersymmetric relics are interesting because of the following
properties: i) if R--parity is conserved, the Lightest Supersymmetric
Particle (LSP) is stable; ii) if colourless and uncharged, the LSP is
a nice realization of a relic WIMP \cite{nic}. In the present paper we
assume that supersymmetry exists in Nature and that properties (i) and
(ii) hold.

The nature of the LSP depends on the susy--breaking mechanism and on
the specific regions of the susy parameter space considered.  Here we
limit ourselves to gravity--mediated schemes, and domains of the
parameter space where the LSP is the neutralino.  The neutralino is
defined as the lowest--mass linear superposition of photino ($\tilde
\gamma$), zino ($\tilde Z$) and the two higgsino states ($\tilde
H_1^{\circ}$, $\tilde H_2^{\circ}$): $\chi \equiv a_1 \tilde \gamma +
a_2 \tilde Z + a_3 \tilde H_1^{\circ} + a_4 \tilde H_2^{\circ}$.
Hereafter, the nature of the neutralino is classified in terms of a
parameter $P$, defined as $P \equiv a_1^2 + a_2^2$.  The neutralino is
called a gaugino when $P > 0.9$, a higgsino when $P < 0.1$, mixed
otherwise.

Extensive calculations on relic neutralino phenomenology in
gravity--mediated models have been performed (among the most recent
references on neutralino dark matter see, for instance,
\cite{noi,comp,probing,bere1,altri,acc,cn,gabr,fcomune}).

Even limiting oneself to a the Minimal Supersymmetric extension of the
Standard Model (MSSM), a significant variety of different schemes
exists, from those based on universal or non-universal supergravity,
with susy parameters defined at the grand unification scale (GUT), to
an effective supersymmetric model defined at the Electro--Weak (EW)
scale.

\subsection{Universal and non--universal SUGRA}

  The MSSM is described
by a Yang--Mills Lagrangian, by the superpotential, which contains all
the Yukawa interactions between the standard and supersymmetric
fields, and by the soft--breaking Lagrangian, which models the
breaking of supersymmetry.  The
Yukawa interactions are described by the parameters $h^i$, which
are related to the masses of the standard fermions by the usual
expressions, {\em e.g.}, $m_t = h^t v_2, m_b = h^b v_1$, where 
$v_1$ and $v_2$ 
are the $vev$'s of the two Higgs fields, $H_1$ and $H_2$.

 Implementation of  this model within a supergravity scheme 
 leads naturally to a set of unification assumptions at a Grand
 Unification (GUT) scale, $M_{GUT}$:  
 i) Unification  of the gaugino masses:
        $M_i(M_{GUT}) \equiv m_{1/2}$,
  ii) Universality of the scalar masses with a common mass denoted by
     $m_0$: $m_i(M_{GUT}) \equiv m_0$, iii) Universality of the
     trilinear scalar couplings:
         $A^{l}(M_{GUT}) = A^{d}(M_{GUT}) = A^{u}(M_{GUT}) \equiv A_0 m_0$. 
      The
    relevant parameters of the model at the electro--weak (EW) scale are
    obtained from their corresponding values at the $M_{GUT}$ scale by running
    these down according to the renormalization group equations (RGE). By
    requiring that the electroweak symmetry breaking is induced radiatively by
    the soft supersymmetry breaking, one finally reduces the model parameters
    to five: $m_{1/2}, m_0, A_0, \tan \beta (\equiv v_2/v_1)$ and sign $\mu$.
 In this very strict scheme (universal SUGRA) the phenomenology of
    relic neutralinos is very sensitive to the way in which various
    constraints (for instance those on the bottom quark mass, $m_b$,
    on the   
    top quark mass, $m_t$, and on the strong coupling
    $\alpha_s$) are implemented.

Models with unification conditions at the GUT scale
represent an  appealing scenario; however,
some of the assumptions listed above, particularly ii) and iii), are not
very solid, because, as was  already emphasized some time ago \cite{com},
universality might occur at a scale higher than $M_{GUT}\sim 10^{16}$
GeV, {\em e.g.}, at the Planck scale. 

To take  into account the uncertainty in the
 unification scale one may introduce deviations in the
unification conditions at $M_{GUT}$.     For instance, deviations from 
universality in the scalar  masses at  $M_{GUT}$, which split 
$M_{H_1}$ from $M_{H_2}$ may be parametrized as 
$M_{H_i}^2 (M_{GUT}) = m_0^2(1 + \delta_i)$. 
This is the case of non--universal SUGRA (nuSUGRA) that we considered 
 in Refs. \cite{comp,probing,bere1}. 
Further extensions of deviations from universality in SUGRA models
 which  include squark and/or gaugino masses are discussed, for instance,
 in \cite{acc,cn}.

The possibility that
 the initial scale for the RGE running, $M_I$, might be smaller than 
 $M_{GUT}\sim 10^{16}$ GeV has been raised,  on the basis of
 a number of string models (see, for instance, 
\cite{gabr,iban,abel} and references quoted therein). As is stressed in 
Ref.\cite{iban}, $M_I$ might be anywhere between the EW
scale and the Planck scale, with significant consequences for
 neutralino phenomenology, {\it e. g.}  for the size of
the neutralino--nucleon cross section.

\subsection{Effective MSSM}

In view of 
 the large uncertainties involved in the choice of the scale $M_I$,  
 the SUGRA schemes turn out to be somewhat problematic: 
the originally appealing feature of a universal SUGRA with few parameters 
fails, because of the need to take into consideration the variability of $M_I$ 
or, alternatively, to add new parameters which quantify the various 
 effects of deviation from universality at the GUT scale. Thus, it
appears  more convenient  to work  with a phenomenological 
susy model whose  parameters are defined directly at the electroweak 
scale. Here, this effective scheme of MSSM is denoted as effMSSM. It   
 provides, at the EW scale, a model, defined in terms of a minimum  number of 
parameters: only those necessary to shape the essentials of the theoretical 
structure of MSSM, and of its particle content. 
Once all experimental and theoretical constraints are implemented in
this effMSSM model, one may investigate its compatibility with
specific theoretical schemes at the desired $M_I$.

In the effMSSM scheme we consider here, we impose  a set of
assumptions at the electroweak scale: 
a) all trilinear parameters are set to zero except those of the third family, 
which are unified to a common value $A$;
b) all squark  soft--mass parameters are taken  
degenerate: $m_{\tilde q_i} \equiv m_{\tilde q}$; 
c) all slepton  soft--mass parameters are taken  
degenerate: $m_{\tilde l_i} \equiv m_{\tilde l}$; 
d) the $U(1)$ and $SU(2)$ gaugino masses, $M_1$ and $M_2$, are 
assumed to be linked by the usual relation 
$M_1= (5/3) \tan^2 \theta_W M_2$ (this is the only GUT--induced
relation we are using, since gaugino mass unification appears to be
better motivated than scalar masses universality). 
As a consequence, the supersymmetric 
parameter space consists of seven independent parameters. 
We choose them to be: 
$M_2, \mu, \tan\beta, m_A, m_{\tilde q}, m_{\tilde l}, A$ 
($m_A$ is the mass of
the CP-odd neutral Higgs boson).  

Much larger extensions of the
supersymmetric models could be envisaged: for
instance,   non--unification of the gaugino masses \cite{cn,griest},
and schemes with CP--violating phases \cite{cp}.

In the present note we only report results obtained in the framkework
 of the effMSSM. For a comparison with results derived in other susy
 schemes see Ref.\cite{probing}.

\subsection{Constraints on supersymmetric parameters}

In our exploration of the susy parameter space, we have implemented
 the following experimental constraints:  
 accelerators data on supersymmetric
and Higgs boson searches (CERN $e^+ e^-$ collider LEP2 \cite{LEPb} 
and Collider
Detector CDF at Fermilab  \cite{cdf}); measurements of the 
$b \rightarrow s + \gamma$ decay \cite{bsgamma}.

A further constraint is supplied by the bounds on the 
supersymmetric  contribution to muon anomalous magnetic moment
 $a_{\mu}^{susy}$. 
The Muon (g-2) Collaboration at Brookhaven National Laboratory has
recently reported a new experimental determination of 
$a_{\mu} \equiv (g-2)/2$ \cite{anom};  this confirms their previous
measurement \cite{anomalous} with a precision increased by a factor of
two. Unfortunately, the theoretical evaluation of the SM contribution
to $a_{\mu}$, $a_{\mu}^{SM}$, is affected by an error which is about
twice the present experimental uncertainty. If we use for the leading
hadronic contribution the average of the evaluations in 
Refs.\cite{dh,j,n} (see also Ref.\cite{dty}), for higher order
corrections the results of Refs.\cite{k} and for the hadronic light--by--light
contribution the average of the results of Refs.\cite{kn,hk}, we find
that $a_{\mu}^{SM}$ deviates from $a_{\mu}^{exp}$ \cite{anom} only
by a mere 1.4$\sigma$, {\it i.e.} that 
$\Delta a_{\mu} \equiv a_{\mu}^{exp} - a_{\mu}^{SM}$ sits in the 
rounded--off 
2$\sigma$ range: $-100 \leq \Delta a_{\mu} \cdot 10^{11} \leq 550$. 
In the present paper we restrain $a_{\mu}^{susy}$ to stay inside this
range. Parenthetically, we note that $\Delta a_{\mu}$ would rise to a 
2.6$\sigma$ level, if for the leading hadronic contribution one 
arbitrarily uses only the results of Ref.\cite{dh} and disregards
those of Refs.\cite{j,n,dty}.

\section{Detection rates for relic neutralinos}

Relic neutralinos may produce many different effects, 
generated either by their interactions with ordinary matter  
or by their pair--annihilations. These effects may be detected in 
experiments designed to search for WIMPs, either directly or
indirectly.  

\subsection{Direct detection} 

Direct detection aims at the measurement of the energy released by a
WIMP to a target nucleus as a consequence of their mutual interaction
\cite{morales}.  The detection rate is proportional to the
WIMP--nucleus cross--section convoluted over the phase--space
distribution function of the WIMPs in the galactic halo, calculated at
the solar--system neighbourhood.  As is customary, we assume that this
distribution function may be written as the product of a matter density
$\rho_{\chi}$ times a velocity distribution funcion $f(\vec{v})$.
Once a specific $f(\vec{v})$ is chosen, measurements of the detection
rate in direct WIMP experiments provide information on the product of
$\rho_{\chi}$ times the WIMP--nucleus cross--section, or on the
product $\rho_{\chi} \; \sigma_{scalar}^{(nucleon)}$, when coherent
interactions dominate over spin--dependent ones
($\sigma_{scalar}^{(nucleon)}$ denotes the WIMP--nucleon scalar
cross--section) \cite{bdmsbi}. We further express $\rho_{\chi}$ as
$\rho_{\chi} = \xi \cdot \rho_l$, where $\rho_l$ is the local {\it
total} dark matter density and $\xi$ ($\xi \leq 1$) is a scaling
parameter which accounts for the actual fraction of local dark matter
to be ascribed to the candidate $\chi$.  Thus, in the case of WIMPs
with coherent interactions with matter, the range of WIMP parameters
under current exploration in WIMP direct searches may approximately be
given as:
\begin{equation}
10^{-10} \; {\rm nbarn} \lsim \
\xi \sigma_{scalar}^{(nucleon)} \lsim 
 2 \cdot 10^{-8} \; {\rm nbarn} \, ,
\label{eq:sens}
\end{equation}
for WIMP masses $m_{\chi}$ in the range:
\begin{equation}
30 \; {\rm GeV} \lsim  m_\chi \lsim 270 \;  {\rm GeV}. 
\label{eq:mass}
\end{equation}      

In deriving the ranges in
Eqs. (\ref{eq:sens}--\ref{eq:mass}),  a
variety of WIMP galactic distribution functions  and
uncertainties in astrophysical 
quantities  have been taken into account \cite{belli1,belli2}. 
When interpreted as due to a WIMP with coherent
interactions, the DAMA annual--modulation effect \cite{dama} provides a 
$3\sigma$ region in the plane $m_{\chi} -  \xi
\sigma_{scalar}^{(nucleon)}$ which is 
embedded in the range of Eqs.(\ref{eq:sens}--\ref{eq:mass}). Other 
experiments of WIMP direct detection  provide upper bounds on 
$\xi \sigma_{scalar}^{(nucleon)}$
\cite{morales,cdms,edel}. It is worth stressing  that a comparison of results
derived from experiments  using different target nuclei is only
feasible under the assumption that the WIMP--nucleus interaction 
is dominantly coherent; otherwise, the comparison is either impossible
or strongly model--dependent. Interpretation of the DAMA
annual--modulation results in terms of a WIMP with a mixed (coherent
and spin dependent) coupling is discussed in Ref. \cite{mixed}.  

\subsection{Indirect detection} 

 Indirect signals may be 
produced by: 

a) Annihilation of neutralinos inside 
 celestial bodies (Earth and Sun), where $\chi$'s have been 
captured and accumulated.  Among the various annihilation 
products, neutrinos would be able to emerge form the celestial bodies 
and be detectable as up--going muons in neutrino telescopes. 

b) Neutralino--neutralino annihilations taking place in the 
galactic halo.  Specific signals would emerge: most 
notably, neutrino fluxes, photon fluxes, and rare components 
in cosmic rays (positrons, antiprotons, antideuterons). 

In case of signals of category (a) the detection rate is, as in the
case of direct detection, proportional to the product $\rho_{\chi} \;
\sigma_{scalar}^{(nucleon)}$ (assuming that coherent interactions
dominate over spin--dependent ones); for signals of category (b), the
detection rates are proportional to the product $\bar\rho_{\chi}^2 \;
<\sigma_{ann} v>$, where $\bar\rho_{\chi}^2$ is the average of
$\rho_\chi^2(\vec r)$ over the galactic halo. This makes the size of
signals of type (b) very sensitive to effects of clumpiness in spatial
distribution of WIMPs \cite{salati}.  Actually, recent
high--resolution simulations of cosmic structures appear to favour
formation of regions of matter overdensity inside galaxies
\cite{moore}, with important implications for neutralino signals (see,
for instance, \cite{beug}).

One may  wonder whether constraints on neutralinos could be
derived from some other indirect means, not related to specific 
experiments designed to search for WIMPs. For instance, it has been
argued that significant constraints could be deduced from 
helioseismology or from data on solar neutrino fluxes \cite{lopes}. 
However, it has been subsequently proved that no constraints can be derived 
at present from solar physics for WIMPs with masses above 30 GeV 
\cite{bffrsv}.

\section{Results}

\begin{figure} [t]
\includegraphics[scale=0.45, bb = 41 60 516 530,clip] {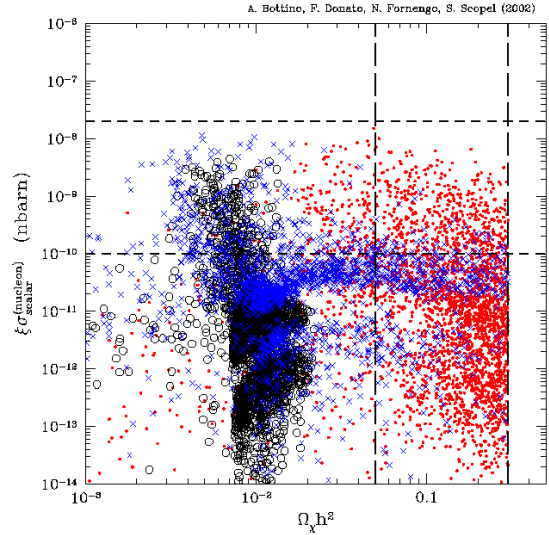}
\vspace{-40pt}
\caption{Scatter plot of $\sigma_{scalar}^{(nucleon)}$ versus 
$\Omega_{\chi}$ in the effective MSSM model described in the text.  
Only configurations with positive $\mu$ and with $m_{\chi}$ in the
range of Eq.(\ref{eq:mass}) are shown. 
The two horizontal lines bracket
the sensitivity region for the WIMP--nucleon scalar cross-section
defined in 
Eq.(\ref{eq:sens}). The
two vertical lines denote the range $0.05 \leq \Omega_{\chi} \leq 0.3$. 
 Dots denote gauginos, circles denote higgsinos and
crosses denote mixed configurations. 
\label{fig:fig1}
}
\end{figure}

\begin{figure} [t]
\includegraphics[scale=0.45, bb = 41 170 516 655,clip] {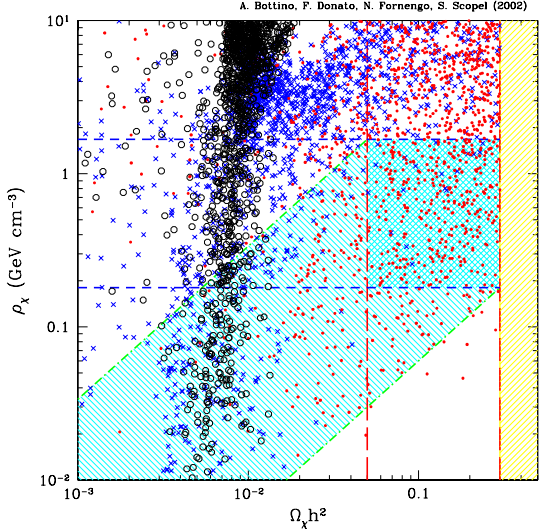}
\vspace{-40pt}
\caption{Scatter plot of $\rho_{\chi}$ versus $\Omega_{\chi}
h^2$. This plot is derived from the value $[\rho_{\chi}$/(0.3 GeV
cm$^{-3}$) $\cdot \sigma^{\rm (nucleon)}_{\rm scalar}]_{expt} = 1
\cdot 10^{-9}$ nbarn and by taking $m_{\chi}$ in the range of
Eq.(\ref{eq:mass}), according to the procedure outlined in the text,
in case of effMSSM. The two horizontal lines delimit the range 0.2 GeV
cm$^{-3}$ $\leq \rho_{\chi} \leq$ 1.7 GeV cm$^{-3}$ \cite{belli2}.
The two vertical lines delimit the range $0.05 \leq \Omega_{\chi} \leq
0.3$.  Dots denote gauginos, circles denote higgsinos and crosses
denote mixed configurations.
\label{fig:fig2}
}
\end{figure}

A first important question is whether present sensitivities in WIMP
direct and indirect experiments are enough to explore some parts of
the supersymmetric parameter space. Furthermore, one wonders whether
the relevant neutralinos provide a significant contribution to
$\Omega_m$. These two questions are connected, since $\Omega_{\chi}
h^2$ and $\sigma_{scalar}^{(nucleon)}$ (and consequently also the
neutralino direct detection rate) are approximately anticorrelated
\cite{lat,cosmo}.

Fig.1 provides an answer to the above questions.  First, it shows that
present experiments of WIMP direct search actually explore significant
regions of the supersymmetric parameter space. This property was
already pointed out in Ref.\cite{bdm}; actually, in the mid--nineties,
with the advent of new detectors, WIMP direct searches acquired the
capability of exploring regions of the susy parameter space not yet
excluded by accelerator bounds. Since then, the progressively more
stringent limits established by new accelerator data have been
counterbalanced by the improved sensitivities of WIMP direct
experiments \cite{noi,comp,probing}. For analyses by other groups see
for instance Refs.\cite{altri,acc,cn,gabr,fcomune}.

Secondly, Fig.1 shows that exploration by WIMP direct search is
effective also in sectors where relic neutralinos may provide a
substantial contribution to $\Omega_m$.  In view of the
anticorrelation between $\Omega_{\chi} h^2$ and
$\sigma_{scalar}^{(nucleon)}$, this property is far from being
trivial.

However, also the case of relic neutralinos which only constitute a
subdominant dark matter population are interesting in view of their
explorability by WIMP direct experiments \cite{lat,cosmo}.  To further
illustrate this point, in Fig.2 we display a scatter plot of the susy
configurations which belong to the value $[\rho_{\chi}$/(0.3 GeV
cm$^{-3}$) $\cdot \sigma^{\rm (nucleon)}_{\rm scalar}]_{expt} = 1
\cdot 10^{-9}$ nbarn, which is representative of a value of
$\rho_{\chi} \cdot \sigma^{\rm (nucleon)}_{\rm scalar}$ inside the
experimental sensitivity range of Eq.(\ref{eq:sens}), and to the mass
interval of Eq.(\ref{eq:mass}). The plot is obtained in the following
way: 1) $\rho_{\chi}$ is evaluated as $[\rho_{\chi} \cdot \sigma^{\rm
(nucleon)}_{\rm scalar}]_{expt}$ / $\sigma^{\rm (nucleon)}_{\rm
scalar}$ ($\sigma^{\rm (nucleon)}_{\rm scalar}$ is calculated within
the susy model), 2) to each value of $\rho_{\chi}$ one associates the
corresponding calculated value of $\Omega_{\chi} h^2$. For a full
discussion of the various sectors displayed in Fig.2 we refer to
\cite{cosmo}. Here we wish to call the attention only to two of these
sectors. The most interesting one is obviously the rectangular
(cross-hatched) region, where 0.2 GeV cm$^{-3} \leq \rho_{\chi} \leq $
1.7 GeV cm$^{-3}$ \cite{belli2} and $0.05 \leq \Omega_{m} h^2 \leq
0.3$; this corresponds to a situation where the neutralino may be a
dominant dark matter component. A second interesting region is the
band delimited by the slanted dot--dashed lines and simply--hatched in
the figure. For configurations which fall inside this band, the
neutralino would provide only a fraction of the cold dark matter, at
the level of local density and of the average relic abundance; a
situation which would be possible, for instance, if the neutralino is
not the unique cold dark matter particle component. To neutralinos
belonging to these configurations one should assign a {\it rescaled}
local density. Our results prove that also these relic neutralinos are
actually probed by WIMP direct searches.

We turn now to indirect WIMP searches. It may be shown that also
measurements of up--going muons at neutrino telescopes
\cite{baksan,macro,sk,amanda} might be capable of exploring
interesting neutralino configurations \cite{neutelesc}, with some
overlap with those analysed in WIMP direct detection experiments
\cite{ntel}.  Fig.3 displays the scatter plot for the up-going muon
flux from the center of the Earth, evaluated in effMSSM, for a
standard galactic distribution function and taking into account the
effect of $\nu_\mu \rightarrow \nu_\tau$ oscillations, for neutrino
mass and mixing angle parameters which best fit the atmospheric
neutrino data \cite{atm,globalfit}.

 A number of susy configurations are at the level of current
experimental bounds. However, one should be aware of the fact that the
efficiency of capture of WIMPs by the Earth, and then also the
up-going muon flux from the center of the Earth, are very sensitive to
the modifications induced by the Sun on the local low-energy WIMP
population.  Depending on the Damour-Krauss conjecture \cite{dk} or
the Gould-Alam conjecture \cite{ga}, the up-going muon flux from the
center of the Earth might be either enhanced or suppressed, as
compared to the standard one. Therefore, at present, derivations of
constraints on supersymmetric configurations from experimental bounds
on up-going muon flux from the center of the Earth is very
hazardous. This important point is often disregarded in the
literature.

\begin{figure} [t]
\includegraphics[scale=0.45, bb = 41 170 516 655,clip] {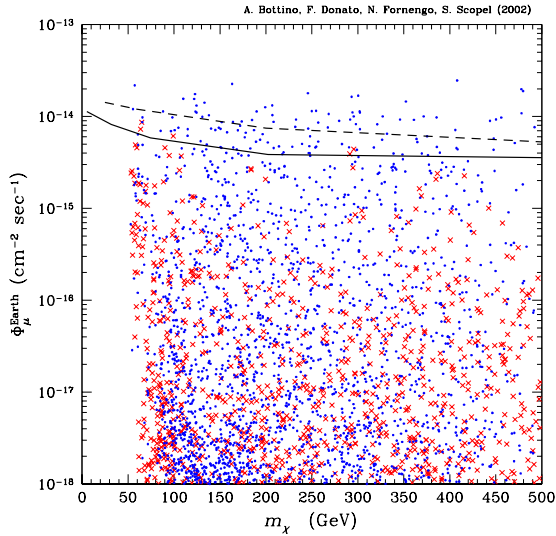}
\vspace{-55pt}
\caption{Scatter plot for the up-going muon flux due to
neutralino-neutralino annihilation in the center of the Earth in
effMSSM.  Dots (crosses) denote configurations with $\Omega_{\chi} h^2
< 0.05$ ($0.05 \leq \Omega_{\chi} h^2 \leq 0.3$). The dashed line
denotes the 90 \% C.L. upper bound of Ref.\cite{macro}, the solid line
denotes the 90 \% C.L. upper bound of Ref.\cite{sk}.
\label{fig:fig1c}
}
\end{figure}

\begin{figure} [t]
\includegraphics[scale=0.45, bb = 41 170 516 655,clip] {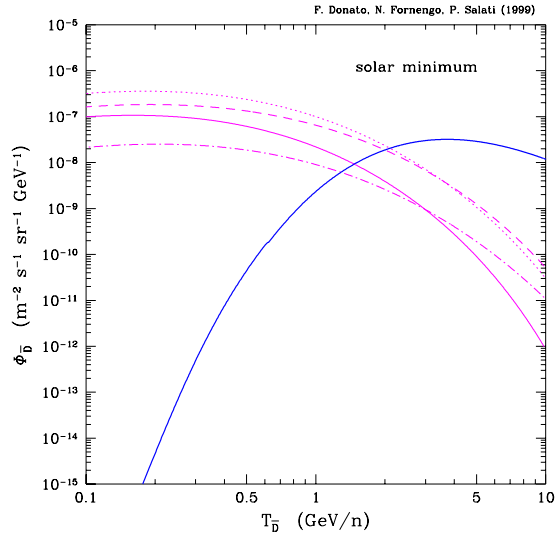}
\vspace{-55pt}
\caption{Top--of--atmosphere antideuteron energy spectra from
Ref. \cite{antid}. The solid line displays the secondary flux, the
other lines denote primary fluxes for four different neutralino
compositions (see Ref. \cite{antid} for details).
\label{fig:fig1c}
}
\end{figure}

Very promising are also strategies for measuring in space signals due
to neutralino pair--annihilations taking place  in the halo. 
Here we briefly mention only some signals consisting of  rare
components in cosmic rays. 

One of the most widely discussed case is
a possible primary flux of low-energy antiprotons. 
Calculations of secondary antiprotons, due to interactions of
cosmic rays with interstellar material, have greatly improved in
recent years \cite{pbar,antip} and have shown that 
at low-energy (antiproton kinetic
energy $T_{\bar p}$ less then about 1 GeV) the secondary spectrum is
much flatter than previously believed and fits remarkably well
the experimental data \cite{bess}. Therefore, only tiny room is left
to exotic components, and the discrimination between secondary
and primary components becomes very problematic. However, since a number of
supersymmetric configurations reach or even exceed the level of the 
experimental spectrum \cite{pbar}, high statistics in $\bar p$ 
measurements and 
great accuracy in the evaluation of the primary and secondary spectra 
may provide useful constraints for supersymmetry. 

Recently, it has been suggested that a measurement of 
 low--energy antideuterons in space   would provide a way
of looking at a distinctive signature of WIMP pair annihilation in the
halo \cite{antid}. In fact at energies below about 1 GeV per nucleon
 the primary
spectrum would be quite dominant over the secondary one (see Fig. 4). 
This property is at variance with what happens in case of the
antiproton low--energy spectra. Thus, the antideuteron measurements open 
up quite new interesting perspectives for detecting $\chi - \chi$ 
annihilation in space.

We wish also to mention that the Heat Collaboration
\cite{heat} reported evidence for an excess of positrons in cosmic
rays at energies around 10 GeV. This might be considered as due to 
neutralino--neutralino annihilation (see for instance \cite{pos}). 
However, it should be noted that this interpretation 
requires an {\it ad hoc} boost factor in order to fit  the
experimental data. 

Other important signals possibly produced by 
neutralino pair--annihilations  in the halo are  provided by
gamma-rays; for this class of processes see, for instance, 
Refs.\cite{beu,bm}. 

In conclusion, it may be said that quite different strategies
of searching for WIMPs are now capable of  
 providing independent information on supersymmetric relic particles.

\end{document}